\documentclass[twocolumn]{revtex4-2}
\usepackage{amsmath}
\usepackage{amssymb,bm}
\usepackage{graphicx}
\usepackage{epstopdf}
\usepackage{hyperref}
\usepackage{bbm}
\usepackage{xcolor}
\usepackage{setspace,comment}
\usepackage{wrapfig}
\newcommand{\eq}[1]{Eq. (\ref{#1})}

\allowdisplaybreaks

\begin{document}
\title{Tunable plasmon-enhanced second-order optical nonlinearity in transition-metal-dichalcogenide nanotriangles}

\author{F. Karimi}\email{farhadk@gmail.com}
\author{S. Soleimanikahnoj}\email{si.soleimani@gmail.com}
\author{I. Knezevic}\email{iknezevic@wisc.edu}

\affiliation{Department of Electrical and Computer Engineering, University of Wisconsin-Madison, Madison, Wisconsin 53706, USA}

\begin{abstract}
The development of nanomaterials with a large nonlinear susceptibility is essential for nonlinear nanophotonics. We show that transition-metal-dichalcogenide (TMD) nanotriangles have a large effective second-order susceptibility [$\chi^{(2)}$] at mid-infrared to near-infrared frequencies owing to their broken centrosymmetry. $\chi^{(2)}$ is calculated within the density-matrix formalism that accounts for dissipation and screening. $\chi^{(2)}$ peaks in the vicinity of both two-photon resonances (specified by the geometry) and plasmon resonances (tunable via the carrier density). Aligning the resonances yields the values of $\chi^{(2)}$ as high as $10^{-6}$ m/V. These findings underscore the potential of TMD nanotriangles for nonlinear nanophotonics, particularly second-harmonic generation.
\end{abstract}
\maketitle

Nonlinear frequency conversion processes, such as second-harmonic generation (SHG) and third-harmonic generation, have many applications in nanophotonics~\cite{boyd2008nonlinear, sun2016optical,sun2018electrically,wang2018nanophotonic,krasnok2018nonlinear,smirnova2016multipolar}. These phenomena rely on intrinsically weak matter-mediated photon--photon interactions ~\cite{kauranen2012nonlinear,khurgin2014graphene,cox2014electrically}, which can can be enhanced through a number of techniques, such as dielectric confinement through Mie resonances~\cite{Grinblat2016_NanoLett_Mie, Camacho-Morales2016_NanoLett_Mie}, quantum confinement~\cite{Miller1985_PRB,Cooperman1982_APL,Yuen1983_APL,Ahn1987_JQE,Khurgin1988_PRB,Khurgin89_JOSA_B}, and surface-plasmonic field enhancement~\cite{cox2014electrically,cox2017quantum,cox2017plasmon,cox2015plasmon,karimi2017plasmons,mikhailov2011theory,karimi2018nonlinear,mesch2016nonlinear}.

The nonlinear and plasmonic response of two-dimensional materials, such as graphene~\cite{hendry2010coherent,gu2012regenerative,wu2011purely,zhang2012z,kumar2013third,hong2013optical,vermeulen2016negative,mikhailov2007non,mikhailov2016quantum,rostami2016theory,rostami2017theory,cheng2014third,cheng2015third,margulis2017quadratic,margulis2016electric,karimi2016dielectric} and transition-metal dichalcogenides (TMDs)~\cite{li2016multimodal,woodward2016characterization,wang2013third,Saynatjoki2016_NatCommTriginalWarping,saynatjoki2013rapid,li2013probing}, has been attracting interest for nonlinear optical applications~\cite{autere2018nonlinear,Wen2019_Infomat}. In TMDs with an odd number of layers, there is weak SHG owing to the material's broken centrosymmetry~\cite{Saynatjoki2016_NatCommTriginalWarping,li2013probing}. In recent years, excitonic effects in TMDs~ \cite{Wang2018_RMP_excitonsTMDs} have been shown to greatly amplify SHG~\cite{Wang2015_PRL_GiantSHG_WSe2,Lee2020_PRB_excitons_metamat_platform,Taghizadeh2019_NonlinSelectionRulesTMD}.

It is challenging to achieve plasmonic-field enhancement in two-dimensional materials because the plasmon wave vector far exceeds that of light at the same frequency. One solution is to lower the system dimensionality, from two to zero. Indeed, quasi-zero-dimensional structures support standing plasmonic resonances that can be easily excited and have been shown to yield an enhanced nonlinear response of graphene nanoislands and nanotriangles~\cite{cox2014electrically,cox2015plasmon,cox2017quantum,cox2017plasmon}. It is important to note that noncentrosymmetric shapes, such as triangles, aid SHG~\cite{Khurgin1988_PRB,Khurgin89_JOSA_B}.

%
%
%
%
%

In this Letter, we show that equilateral nanotriangles made of single-layer TMDs such as MoS$_2$, WS$_2$, and WSe$_2$, whose growth has already been demonstrated~\cite{xie2018coherent,wang2018revealing,lauritsen2004atomic,tuxen2010size,helveg2000atomic}, have a strong and electrically tunable second-order nonlinear optical response at midinfrared (mid-IR) to near-infrared (near-IR) frequencies. We calculate \cite{Code} the second-order nonlinear optical response of these systems within the density-matrix framework that accounts for screening and disspation~\cite{boyd2008nonlinear} (excitonic effects are not considered). We show that the second-order susceptibility peaks in the vicinity of both two-photon intersubband resonances (whose positions are fixed by the nanotriangle geometry) and plasmon resonances (dynamically tunable by the carrier density). By tuning the carrier density to bring the plasmon and two-photon resonances into alignment, second-order susceptibility $\chi^{(2)}$ can become as high as 10$^{-6}$ m/V, orders of magnitude higher than the intrinsic SHG of single-layer TMDs ($\sim$ 10$^{-9}$ m/V) or the second-order susceptibility of bulk LiNbO$_3$ ($\sim$ 10$^{-11}$ m/V) at near-IR  frequencies~\cite{schiek2012absolute,Wen2019_Infomat}. The second-order optical response increases as the triangle size decreases. These findings underscore the suitability of single-layer TMD nanotriangles as elements for nonlinear nanophotonics.


\begin{figure*}
   \includegraphics[width=\textwidth]{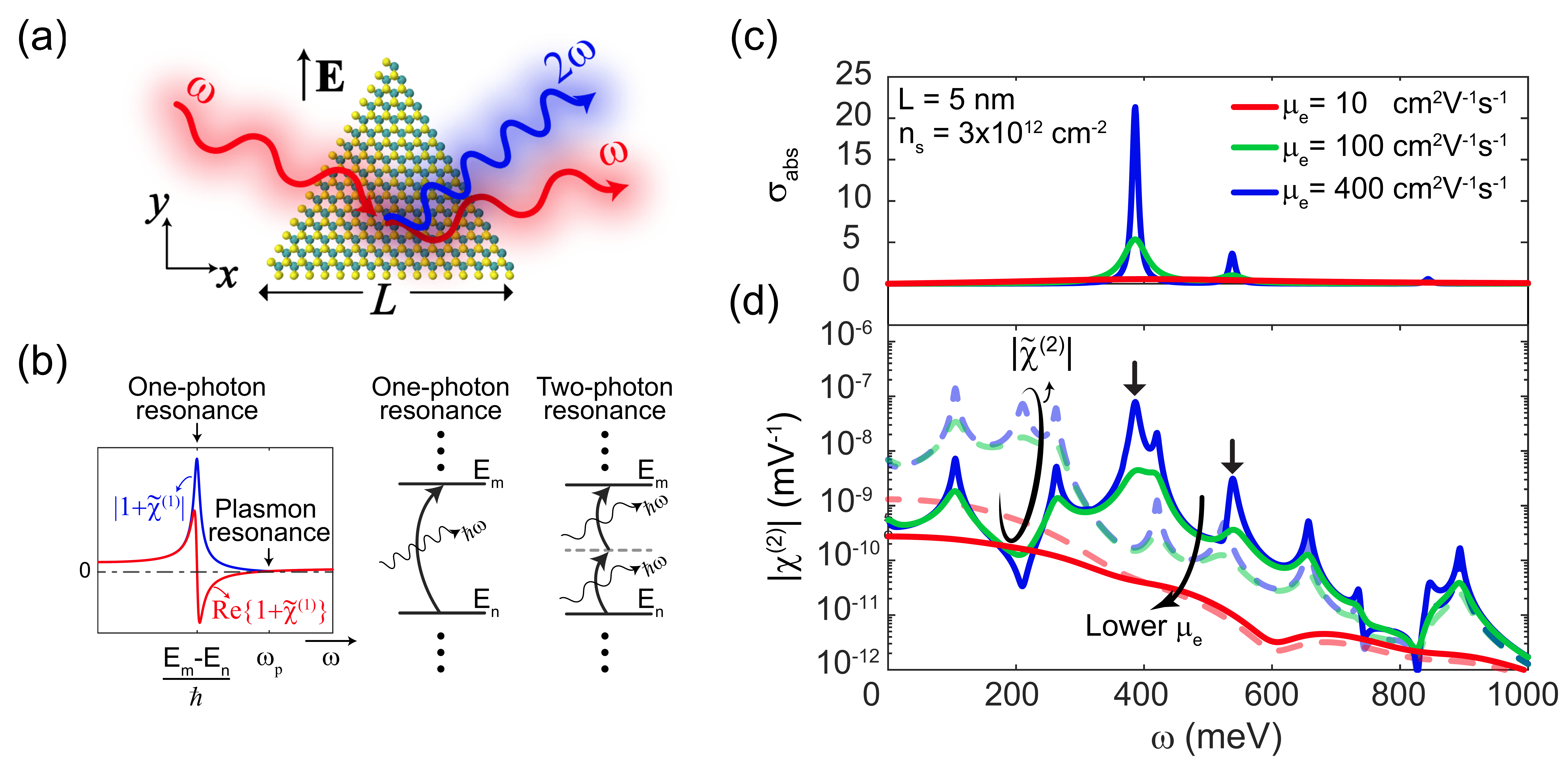}
    \caption{\label{fig1}(a) Schematic of an equilateral TMD nanotriangle with the side length $L$. TM-polarized incident light causes linear as well as second-order nonlinear optical response. (b) One-photon, two-photon, and plasmon resonances. One- and two-photon resonances are due to intersubband optical transitions, whereas plasmon resonances are due to collective oscillations of electrons. (c) The loss function and (d) second-order susceptibility in response to external field (solid) and total field (dashed) of a 5-nm TMD nanotriangle with the sheet density $n_s = 3\times 10^{12}$ cm$^{-2}$ for different values of the electron mobility. The peaks marked by arrows correspond to two different plasmon resonances.}
\end{figure*}

\emph{Density-Matrix Calculation.} TMDs have the general formula of $MX_2$, with $M$ and $X$ representing a transition metal and a chalcogen, respectively. Single-layer $MX_2$ ($M$=Mo, W; $X$=S, Se) are direct-bandgap semiconductors~\cite{rasmussen2015computational}. Since these TMDs have very similar electron effective masses (between 0.46 and 0.55 $m_e$), same-size nanotriangles of them are expected to have similar optical responses. Therefore, in this paper, we only study the second-order nonlinear response of MoS$_2$ nanotriangles (effective mass 0.55 $m_e$~ \cite{rasmussen2015computational}).

Using the density-matrix method~\cite{boyd2008nonlinear}, we calculate \cite{Code} the loss function and second-order susceptibility for an MoS$_2$ equilateral nanotriangle with the side length $L$, area $A$, thickness $d$, and sheet carrier density $n_s$. The nanotriangle lies on the $z=0$ plane upon a substrate that fills the $z<0$ half-space and has an absolute dielectric permittivity $\varepsilon_b = \varepsilon_0 \kappa_b$. We assume the nanotriangle is illuminated with TM-polarized light with the electric field along the $y$-direction [Fig. \ref{fig1}(a)]. A similar theoretical approach was successfully employed for graphene nanoflakes~\cite{cox2014electrically,cox2017quantum}.

The Hamiltonian describing electrons in a TMD nanotriangle is $\mathbb{H} = \mathbb{P}^2/2m_e+\mathbb{U}_L + \mathbb{U}_W$. $\mathbb{U}_L$ is the lattice potential and $\mathbb{U}_W$ denotes a mesoscopic-scale potential defining the triangle (zero within and infinite elsewhere). The details of the tight-binding representation of the lattice potential $\mathbb{U}_L$ can be found in Refs.~\cite{Khorasani_2018,Rostami2015,Cappelluti2013}, while the database presented in Ref.~\cite{rasmussen2015computational} provides details on the TMD electronic structure, including effective masses. In this paper, we use the envelope-function approximation (EFA), which, given the size of the nanotriangles of interest, is an accurate approximation. Using the EFA, the energies and eigenstates of the Hamiltonian are, respectively, $\epsilon_{nm,\nu} = \epsilon_\nu + 8\pi^2\hbar^2/(9m_\nu^*L^2) (n^2+m^2+nm)$ and $\langle \bm{r}|nm,\nu\rangle = f_{nm}(\bm{r}) u_\nu(\bm{r}) $. $m_\nu^*$ is the electron effective mass near the extremum of the bulk single-layer TMD's $\nu^{th}$ band, with the extremum energy $\epsilon_\nu$ and the periodic part of the Bloch wave function $u_\nu(\bm{r})$. $f_{nm}(\bm{r})$ is a solution to the Schr\"{o}dinger equation for a particle in an equilateral triangle, which can be obtained analytically, as detailed in Refs. ~\cite{Gaddah_2013,Li_1985_SchroedEqn}. The $f_{nm}(\bm{r})$ eigenfunctions fall into two classes:
\begin{subequations}
\begin{equation}
\begin{split}
    f_{nm}^{(1)} (x,y)& =
    N_1\Big\{\sin[\beta(n+2m)\tilde{x}] \sin[\sqrt{3}\beta n \tilde{y}]\\
    & +  \sin[\beta(m-n)\tilde{x}]\sin[\sqrt{3}\beta(n+m)\tilde{y}] \\
    & - \sin[\beta(2n+m)\tilde{x}] \sin[\sqrt{3}\beta m \tilde{y}]\Big\},\\
\end{split}
\end{equation}
\begin{equation}
\begin{split}
    f_{nm}^{(2)} (x,y)& =
    N_2 \Big\{- \cos[\beta(n+2m)\tilde{x}] \sin[\sqrt{3}\beta n \tilde{y}]  \\
    & + \cos[\beta(m-n)\tilde{x}]\sin[\sqrt{3}\beta(n+m)\tilde{y}]\\
    & - \cos[\beta(2n+m)\tilde{x}] \sin[\sqrt{3}\beta m \tilde{y}]\Big\},\\
\end{split}
\end{equation}
\end{subequations}
where $\beta = 2\pi/3L$, $\tilde{x} = x+L/2$, $\tilde{y} =y+\sqrt{3}L/2 $, and the origin of the Cartesian coordinate system is at the triangle's center. For the eigenfunctions of the first class, $m \ge n > 0$. $N_1 = 2 (16/27)^{1/4}/L$ and $N_1 =2 (4/27)^{1/4}/L$ for $m\neq n$ and $m=n$, respectively. For the eigenfunctions of the second class, $m>n>0$ and $N_2 = 2 (16/27)^{1/4}/L$. It should be noted that the EFA is a valid approximation as long as the number of local maxima of a relevant wave function is much smaller than the number of unit cells in the nanotriangle.

Now, once we know the electronic energies and eigenstates in the TMD nanotriangle, we use the density-matrix method to calculate the linear and second-order nonlinear optical susceptibility~\cite{boyd2008nonlinear,cox2014electrically,cox2017quantum}. The linear optical susceptibility can be obtained from
\begin{equation}\label{Chi1}
\begin{split}
    \tilde{\chi}^{(1)}_{ij}(\omega) &= \frac{n_s}{\varepsilon_0 d} \sum_{\bm{ss'}} (\rho^{(0)}_{\bm{s's'}} - \rho^{(0)}_{\bm{ss}})\frac{\mu_{\bm{ss'}}^j \mu_{\bm{s's}}^i}{\epsilon_{\bm{s}}-\epsilon_{\bm{s'}}-\hbar \omega -i\hbar \gamma},\\
\end{split}
\end{equation}
where $\bm{s}$ denotes a set of quantum numbers $\{n,m,\nu\}$, $ijk$ refer to the Cartesian coordinates, and $\gamma$ represents the relaxation rate of electrons. $\bm{\mu}_{\bm{ss'}}$ is the electric dipole transition moment and equals to $\langle \bm{s}|(-e)\bm{r}|\bm{s'}\rangle $. Since $f(\bm{r})$ varies much slower than $u(\bm{r})$, the expression of $\bm{\mu}$ can be simplified. For interband transitions ($\nu\neq\nu'$), $\langle \bm{s}|(-e)\bm{r}|\bm{s'}\rangle \approx \langle \nu|(-e)\bm{r}|\nu'\rangle $; for intraband transitions, ($\nu = \nu'$), $\langle \bm{s}|(-e)\bm{r}|\bm{s'}\rangle \approx \langle nm|(-e)\bm{r}|n'm'\rangle $. In \eq{Chi1}, $\rho^{(0)}_{\bm{ss'}} = \delta_{\bm{s,s'}} F(\epsilon_{\bm{s}})$ denotes the elements of the unperturbed density matrix. $F$ is the Fermi-Dirac distribution function and satisfies $\sum_{\bm{s}} F(\epsilon_{\bm{s}})= A n_s$. [We neglect the spatial variation of the response functions because there is little spatial variation in the induced charge density. Namely, a triangle is a type of quantum dot, so its envelope functions are standing waves. For the lowest few energy levels that stem from confinement, the wavelengths of the propagating waves that make up the standing waves are on the order of the size of the triangle, which is large on crystalline length scales (see the justification for using the EFA above), meaning that their wave vectors would be very small (near the center of the Brillouin zone of the underlying crystalline lattice). Consequently, the associated charge density and screening strength vary slowly with position. We take this weak spatial variation into account in an average sense through the dipole transition moments between confined states.]

The second-order (nonlinear optical) susceptibility is given by
\begin{align}\label{Chi2}
    \tilde{\chi}^{(2)}_{ijk}(2\omega) &= \frac{n_s}{2 \varepsilon_0 d} \sum_{\bm{s'ss''}}
    (\rho^{(0)}_{\bm{s''s''}} - \rho^{(0)}_{ss}) \times \nonumber\\
    & \hspace{-1cm}\Big[ ~~\frac{\mu_{\bm{s''s'}}^i \mu_{\bm{s's}}^j \mu_{\bm{ss''}}^k }
    {(\epsilon_{\bm{s'}}-\epsilon_{\bm{s''}}-2\hbar \omega -i\hbar \gamma)(\epsilon_{\bm{s}}-\epsilon_{\bm{s''}}-\hbar \omega -i\hbar \gamma)} \nonumber\\
    &\hspace{-1cm} + \frac{\mu_{\bm{s''s'}}^i \mu_{\bm{s's}}^k \mu_{\bm{ss''}}^j }
    {(\epsilon_{\bm{s'}}-\epsilon_{\bm{s''}}-2\hbar \omega -i\hbar \gamma)(\epsilon_{\bm{s}}-\epsilon_{\bm{s''}}-\hbar \omega -i\hbar \gamma)} \nonumber\\
    &\hspace{-1cm} + \frac{\mu_{\bm{s''s'}}^j \mu_{\bm{s's}}^i \mu_{\bm{ss''}}^k }
    {(\epsilon_{\bm{s'}}-\epsilon_{\bm{s}}+2\hbar \omega +i\hbar \gamma)(\epsilon_{\bm{s}}-\epsilon_{\bm{s''}}-\hbar \omega -i\hbar \gamma)} \nonumber\\
    &\hspace{-1cm} + \frac{\mu_{\bm{s''s'}}^k \mu_{\bm{s's}}^i \mu_{\bm{ss''}}^j }
    {(\epsilon_{\bm{s'}}-\epsilon_{\bm{s}}+2\hbar \omega +i\hbar \gamma)(\epsilon_{\bm{s}}-\epsilon_{\bm{s''}}-\hbar \omega -i\hbar \gamma)} \Big].\nonumber\\
\end{align}
Because of the particle-hole symmetry in single-layer TMDs, the contributions from interband transitions in \eq{Chi2} add up to zero. In contrast, intraband optical transitions result in a nonzero $\tilde{\chi}^{(2)}$. Therefore, we pick all eigenstates $\nu$ in the first conduction band of the single-layer TMD, within which the Fermi level lies.

Moreover, owing to the $D_3$ symmetry of equilateral triangles, $ \tilde{\chi}_{ij}^{(1)}$ and $ \tilde{\chi}_{ijk}^{(2)}$ tensors of the TMD nanotriangles can be simplified~\cite{boyd2008nonlinear}. The only nonzero elements of $ \tilde{\chi}^{(1)}$  are $ \tilde{\chi}^{(1)}_{xx}  = \tilde{\chi}^{(1)}_{yy}$ and the only nonzero elements of $ \tilde{\chi}^{(2)}$  are $ \tilde{\chi}^{(1)}_{yyy} =-  \tilde{\chi}^{(1)}_{yxx} = -  \tilde{\chi}^{(1)}_{xxy} = -  \tilde{\chi}^{(1)}_{xyx}$. In this paper, we only calculate $\tilde{\chi}^{(1)}_{yy}$ and $\tilde{\chi}^{(2)}_{yyy}$ and, to simplify the notation, we drop the $y$ indices henceforth.

There are two types of intersubband optical transitions contributing to the second-order optical response: one-photon transitions corresponding to the terms with $\hbar\omega$ in \eq{Chi2} and two-photon transitions corresponding to the terms with $2\hbar\omega$ in \eq{Chi2} [Fig. \ref{fig1}(b)]. The second-order susceptibility peaks when either one-photon or two-photon transitions are nearly resonant. The second-order susceptibility also peaks at the vicinity of the plasmon resonances. Unlike the one- and two-photon resonances, which stem from intersubband optical transitions, plasmon resonances are due to collective electron oscillations. In a confined quantum system with considerable energy-level spacing, like TMD nanotriangles, each one-photon resonance is succeeded by a plasmon resonances, as illustrated in Fig. 1(b). Plasmon resonance are weaker at higher subbands owing to the lower subband carrier densities.  Unfortunately, $\tilde{\chi}^{(2)}$ does not capture plasmon resonances, because it is calculated in response to the total field acting on electrons, which consists of the external field and the induced field. In order to capture the plasmonic effect and understand the optical response to the external field rather than the total field, we introduce two quantities that are relevant in experiment~ \cite{mikhailov2011theory,cox2017quantum,rostami2017theory}:
\begin{equation}
\begin{split}
    \chi^{(2)} &= \tilde{\chi}^{(2)}/(1+\tilde{\chi}^{(1)})^2,\\
    \sigma_{\text{abs}} &= -\text{Im}\{(1+\tilde{\chi}^{(1)})^{-1}\},\\
\end{split}
\end{equation}
$\sigma_{\text{abs}}$ denotes the loss function, which peaks near the plasmon resonances. $\chi^{(2)}$ is also referred to as the second-order susceptibility with respect to the external field. (Note that we have used $\sim$ to denote the susceptibility with respect to the total field.) Unlike $\tilde{\chi}^{(2)}$, $\chi^{(2)}$ does not peak near one-photon resonances and only peaks in the vicinity of plasmon and two-photon resonances. The strength of these resonances are critically dependent on the electron relaxation rate. For small $\gamma$, the resonances are sharp. For large $\gamma$, scattering-induced broadening practically quenches the resonance. In this paper, we use the electron mobility [$\mu_e = e/(m^*\gamma)$] as a proxy for the characteristic relaxation rate $\gamma$ due to phonons and impurities within the density-matrix formalism~\cite{boyd2008nonlinear,boyd2008nonlinear,cox2014electrically,cox2017quantum}.

\emph{Results and discussion.} Figures \ref{fig1}(c) and \ref{fig1}(d) show the loss function ($\sigma_{\text{abs}}$) and the second-order susceptibility in response to the external field ($\chi^{(2)}$) as well as to the total field ($\tilde{\chi}^{(2)}$). For a fixed side length $L$, the optical response of a nanotriangle is dependent on the electron mobility, or, equivalently, the electron relaxation rate. The measured room-temperature electron mobility of bulk single-layer MoS$_2$ is in the range of 0.5--200 $\mathrm{cm^2/Vs}$~ \cite{radisavljevic2011single,kaasbjerg2012phonon,wang2012electronics}. The lower density of electron states in the nanotriangle with respect to bulk leads to reduced scattering rates; this is a well-known feature of low-dimensional systems~\cite{LundstromBook}. To capture this reduction in the scattering rates, we choose an enhanced value of the maximal effective mobility to be 400 $\mathrm{cm^2/Vs}$ (twice the highest value measured in bulk) as a representative of the high-carrier-mobility range. By increasing the electron mobility, the effect of optical resonances on the second-order optical response becomes more pronounced.(At low electron mobilities [Figs. \ref{fig1}(c) and \ref{fig1}(d)], spectral broadening weakens the optical resonances and $\chi^{(2)}$ varies smoothly as a function of frequency, \textit{i.e.}, is devoid of resonances.) Henceforth, we focus only on the 400 $\mathrm{cm^2/Vs}$ mobility results, where the interesting features are most prominent.

As discussed above, $\tilde{\chi}^{(2)}$ (the second-order susceptibility with respect to the total field) peaks in the vicinity of one- and  two-photon  resonances while $\chi^{(2)}$ (the second-order susceptibility with respect to external field) peaks in the vicinity of two-photon and plasmon resonances. The arrow-marked peaks in Fig. \ref{fig1}(c) correspond to the plasmon resonances. The positions of two-photon resonances are dependent on the single-particle energies in the nanotriangle and are fixed for a given nanotriangle size [Fig. \ref{fig1}(c)]. On the other hand, the positions of plasmon resonances depend on electron energies, but also on the carrier density in the nanotriangle. Therefore, we could dynamically tune the plasmon resonances by changing the carrier density (for example, by using back gating), which suggests that, by aligning the plasmon and two-photon resonances, we could amplify the second-order optical susceptibility.

\begin{figure*}
    \includegraphics[width=\textwidth]{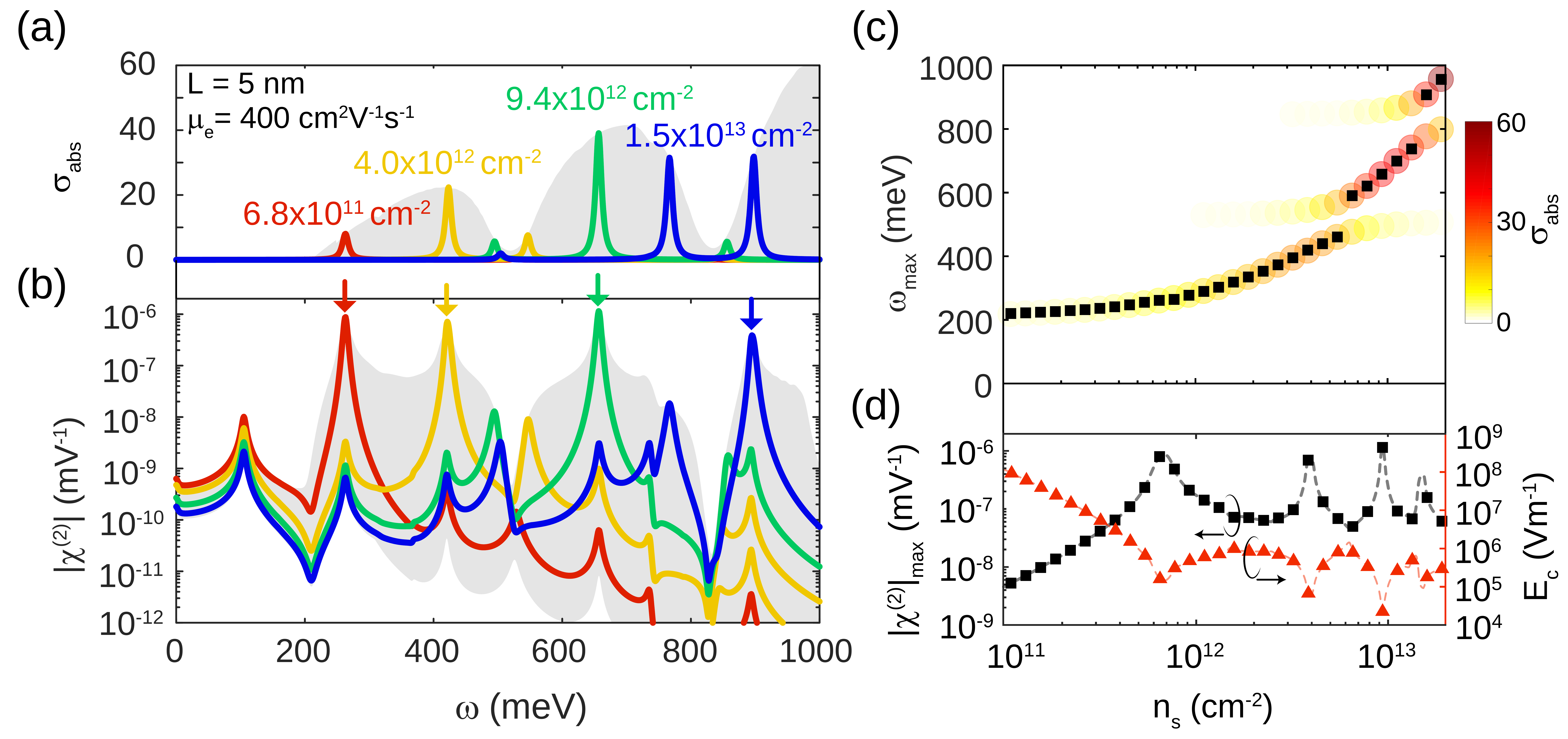}
    \caption{\label{fig2}(a) The loss function and (b) the second-order susceptibility ($\chi^{(2)}$) versus illumination frequency for different carrier densities. The arrow-marked peaks  correspond to plasmon resonances. The gray shaded areas show $\sigma_{\text{abs}}$ and $\chi^{(2)}$ calculated for carrier densities in the $10^{11}$--$3\times 10^{13}$ cm$^{-2}$ range. (c) $\omega_\text{max}$, the frequency of maximum second-order susceptibility (square) and (d) its corresponding values of $|\chi^{(2)}|_\text{max}$ (square) and $E_c$ (triangle) as a function of carrier density. The filled circles in (c) represent plasmon resonances and the color bar shows the corresponding loss function of each plasmon resonance. Note that $\omega_\text{max}$ coincides with a plasmon resonance. In all panels, $L=5$ nm and $\mu_e$ = 400 cm$^{2}$/Vs.}
\end{figure*}

Figures \ref{fig2}(a)--(b) show the loss function $\sigma_{\text{abs}}$ and the second-order susceptibility $\chi^{(2)}$ for different carrier densities. The carrier density at which a plasmon resonance and a two-photon resonance are aligned with each other, $\chi^{(2)}$ increases to as large as $10^{-6}$ m/V. TMD nanotriangles can therefore have almost five orders of magnitude stronger second-order nonlinearity than bulk LiNbO$_3$ at near-IR frequencies ($\sim$  10$^{-11}$ m/V) or the intrinsic second-order response of single-layer SHG ($\sim$  10$^{-9}$ m/V)~ \cite{schiek2012absolute,Wen2019_Infomat}. The gray shaded areas in Figs. \ref{fig2}(a) and \ref{fig2}(b) show $\sigma_{\text{abs}}$ and $\chi^{(2)}$ calculated for the carrier densities in the range of 10$^{11}$--3$\times$10$^{13}$ cm$^{-2}$. Plasmon and one-photon resonances never happen at the same frequency, so there are no-plasmon regions at the vicinity of one-photon resonances [Fig. \ref{fig1}(b)], and, consequently, second-order optical susceptibility $\chi^{(2)}$ decreases in the vicinity of one-photon resonances.

To understand the interplay between these resonances, we define $\omega_{\text{max}}$ as the frequency at which the maximum second-order susceptibility [$|\chi^{(2)}|_{\text{max}}$] happens in a nanotriangle with a given side length, carrier mobility, and carrier density. It can be seen that $\omega_{\text{max}}$ [black squares in Figs. \ref{fig2}(c)--(d)] are in close proximity of a plasmon resonance. In Fig. \ref{fig2}(c), filled circles denote the plasmon resonances and the color bar represents their corresponding loss function. By lowering the carrier density, $\omega_{\text{max}}$ asymptotically decreases to the first one-photon resonance of the nanotriangle. By decreasing the carrier density, the plasmon resonances also weaken and $|\chi^{(2)}|_{\text{max}}$ decreases. Correspondingly, the critical electric field, which is defined as $E_c \equiv {|\chi^{(1)}|_{\text{max}}/|\chi^{(2)}|_{\text{max}}}$ [red triangles in Fig. \ref{fig2}(d)], decreases with increasing carrier density. The lower the critical electric field, the stronger the second-order nonlinear optical response. At high carrier densities, $E_c$ can be as low as $\sim$ 0.1 kVcm$^{-1}$, which corresponds to an optical field intensity as low as 0.1 kW/cm$^2$.

\begin{figure*}
    \includegraphics[width=\textwidth]{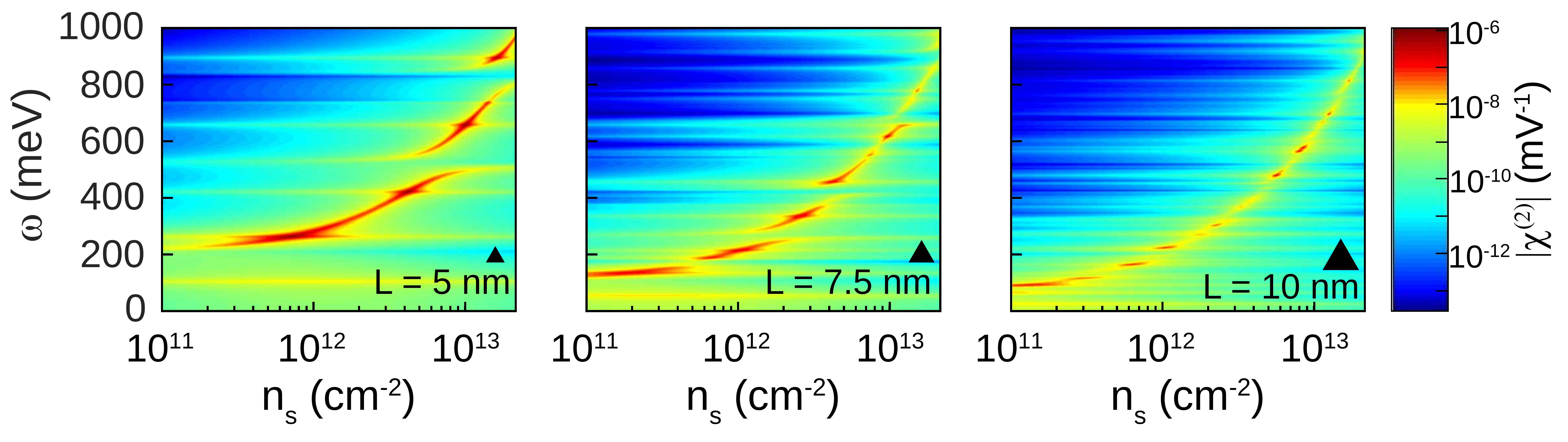}
    \caption{\label{fig3}The magnitude of the second-order susceptibility as a function of frequency and carrier density for an MoS$_2$ nanotriangle with (left) $L=5$ nm, (middle) $L=7.5$ nm, and (right) $L=10$ nm. Smaller nanotriangles exhibit stronger second-order nonlinear optical response. }
\end{figure*}

Finally, we study the effect of the TMD nanotriangle's size on its second-order optical nonlinear response. Figure \ref{fig3} shows $|\chi^{(2)}(\omega)|$ as a function of frequency and carrier density for nanotriangles with different sizes. By increasing the side length $L$, the maximum value of $\chi^{(2)}$ decreases. Increasing $L$ has a twofold effect. On one hand, the magnitude of the dipole moments grows with increasing $L$. On the other hand, the spacing between a nanotriangle's energies drops with increasing $L$ and results in more one- and two-photon processes contributing to the second-order nonlinear response. Since these one- and two-photon processes have different phases, the net effect of these competing phenomena results in the lower $|\chi^{(2)}(\omega)|$ for larger $L$. Also, owing to the smaller energy spacing, the first horizontal asymptote of $\omega_{\text{max}}$ occurs at a lower frequency in large nanotriangles.

\emph{Conclusion.} We showed that, in the mid-IR to near-IR frequency range, TMD nanotriangles have a large and tunable second-order susceptibility, which makes them promising materials for integrated nonlinear-nanophotonics applications. Using the density-matrix method with an account of screening and dissipation, we calculated the linear and second-order optical response of TMD nanotriangles. We showed that second-order susceptibility peaks at the vicinity of plasmon and two-photon resonances. For a given material, the two-photon resonances are fixed by the nanotriangle's size while the plasmon resonances can be tuned via the carrier density. By aligning the plasmon and two-photon resonances, second-order susceptibility can become as high as  $10^{-6}\,\mathrm{m/V}$, with higher magnitudes found in smaller triangles.

If a triangle deviates from perfect $D_3$ symmetry, which is relevant in experiment, the second-order nonlinear susceptibility tensor will have more than four nonzero elements. There will be many more nonzero dipole transition moments, meaning more one- and two-photon processes with different phases, which might reduce the second-order nonlinear susceptibility.

\emph{Acknowledgment.} The authors gratefully acknowledge support by the U.S. Department of Energy, Office of Basic Energy Sciences, Division of Materials Sciences and Engineering, Physical Behavior of Materials Program, under Award DE-SC0008712. Preliminary work was partially supported by the Splinter Professorship. This work was performed using the compute resources and assistance of the UW-Madison Center for High Throughput Computing (CHTC) in the Department of Computer Sciences.


%

\end{document}